\documentclass[%
reprint, superscriptaddress,
 amsmath,amssymb,
 aps,
]{revtex4-1}

\usepackage{graphicx}
\usepackage{dcolumn}
\usepackage{bm}

\begin{document}

\preprint{APS/123-QED}

\title{Probing proximity induced superconductivity in InAs nanowire using built-in barriers}

\author{Tosson Elalaily}
 \affiliation{Department of Physics, Budapest University of Technology and Economics and Nanoelectronics 'Momentum' Research Group of the Hungarian Academy of Sciences, Budafoki ut 8, 1111 Budapest, Hungary\\}
\affiliation{Department of Physics , Faculty of Science, Tanta University, Al-Geish St., 31111 Tanta, Gharbia, Egypt.\\}

\author{Oliv\'er K\"urt\"ossy}
 \affiliation{Department of Physics, Budapest University of Technology and Economics and Nanoelectronics 'Momentum' Research Group of the Hungarian Academy of Sciences, Budafoki ut 8, 1111 Budapest, Hungary\\}

\author{Valentina Zannier}
 \affiliation{NEST, Istituto Nanoscienze-CNR and Scuola Normale Superiore, Piazza San Silvestro 12, I-56127 Pisa, Italy\\}

\author{Zolt\'an Scher\"ubl}
 \affiliation{Department of Physics, Budapest University of Technology and Economics and Nanoelectronics 'Momentum' Research Group of the Hungarian Academy of Sciences, Budafoki ut 8, 1111 Budapest, Hungary\\}

\author{Istv\'an Endre Luk\'acs}
 \affiliation{Center for Energy Research, Institute of Technical Physics and Material Science, Konkoly-Thege Mikl\'os \'ut 29-33., H-1121, Budapest, Hungary\\}

\author{Pawan Srivastava}
 \affiliation{Department of Physics, Budapest University of Technology and Economics and Nanoelectronics 'Momentum' Research Group of the Hungarian Academy of Sciences, Budafoki ut 8, 1111 Budapest, Hungary\\}

\author{Francesca Rossi}
 \affiliation{IMEM-CNR, Parco Area delle Scienze 37/A, I-43124 Parma, Italy\\}

\author{Lucia Sorba}
 \affiliation{NEST, Istituto Nanoscienze-CNR and Scuola Normale Superiore, Piazza San Silvestro 12, I-56127 Pisa, Italy\\}

\author{Szabolcs Csonka}
 \email{szabolcs.csonka@mono.eik.bme.hu}
 \affiliation{Department of Physics, Budapest University of Technology and Economics and Nanoelectronics 'Momentum' Research Group of the Hungarian Academy of Sciences, Budafoki ut 8, 1111 Budapest, Hungary\\}

\author{P\'eter Makk}
 \email{peter.makk@mail.bme.hu}
 \affiliation{Department of Physics, Budapest University of Technology and Economics and Nanoelectronics 'Momentum' Research Group of the Hungarian Academy of Sciences, Budafoki ut 8, 1111 Budapest, Hungary\\}

\date{\today}

\begin{abstract}
Bound states in superconductor-nanowire hybrid devices play a central role, carrying information on the ground states properties (Shiba or Andreev states) or on the topological properties of the system (Majorana states). The spectroscopy of such bound states relies on the formation of well-defined tunnel barriers, usually defined by gate electrodes, which results in smooth tunnel barriers. Here we used thin InP segments embedded into InAs nanowire during the growth process to form a sharp built-in tunnel barrier. Gate dependence and thermal activation measurements have been used to confirm the presence and estimate the height of this barrier. By coupling these wires to superconducting electrodes we have investigated the gate voltage dependence of the induced gap in the nanowire segment, which we could understand using a simple model based on Andreev bound states. Our results show that these built-in barriers are  promising  as future spectroscopic tools.
\end{abstract}

\maketitle


\section{Introduction}
\label{sec:Introduction}

Semiconducting nanowires \cite{Lieber2007} recently became one of the leading platforms for nanoscale hybrid devices. The reason for this lies in the high quality materials available and also in the possibility to confine electrons using electrostatic gating.  Therefore these wires became a host  for several fascinating systems: spin-qubits\cite{nadj2010spin}, Andreev qubits\cite{hays2018direct,tosi2019spin}, Cooper pair-splitter\cite{hofstetter2009cooper,das2012high,fulop2015magnetic,fulop2014local} devices, magnetic Weyl-points or the famous Majorana wires\cite{mourik2012signatures,deng2016majorana,zhang2018quantized,lutchyn2018majorana}. Particularly, InAs wires are very widely used due to the large spin orbit coupling, the simplicity of realizing electrical contacts  and the  precise control of growth conditions that can be achieved.
For most of these systems mentioned above superconducting (SC) correlations are needed to be induced in the wires, which is achieved by coupling the wire to SC electrodes. This coupling results in proximity induced superconductivity and the appearance of bound states in the wire. Whereas these bound states can come in different flavour from Andreev\cite{hays2018direct,tosi2019spin,lee2014spin,sand2007kondo}, Shiba\cite{jellinggaard2016tuning,grove2018yu,scherubl2019large,chang2013tunneling} to even Majorana \cite{mourik2012signatures,deng2016majorana,zhang2018quantized,lutchyn2018majorana}, their observation requires the formation of tunnel probes in the vicinity of these states.

Usually, to define tunnel barriers electrostatic gates placed next to or below the wires are used \cite{Fasth2005, Pfund2007}. Whereas these gates allow some tunability of the coupling, the barriers formed are rather smooth and wide, due to the distance of the gate electrode from the wire and the finite width of the gate and wire.  Moreover, such gates modify the potential profile in an extended region of the wire changing also the bound states that one wants to probe. However, well defined, sharp barriers can be formed by embedding InP segments into the InAs wire, during the growth procedure. The barrier originates from the different band alignment of the conduction band of InAs and InP, as shown in Fig.~\ref{fig:device}b. Previous studies have shown that the barriers are atomically sharp, and their width can be controlled with the precision of a single atomic layer\cite{zannier2017nanoparticle}. Using two of these barriers quantum dots were formed \cite{Samuelson2004}, which were used to study coupling asymmetries in a dot \cite{thomas2019highly}, the  Stark effect \cite{Roddaro2011,Rossella2014,Romeo2012} or thermal transport in quantum dots \cite{fuhrer2007few,bleszynski2008imaging,Josefsson2018}.  However, no studies have investigated how well these barriers can be used as a spectroscopy tool for superconducting bound states.

Here we investigate the proximity effect in InAs wires using single InP segments as tunnel barriers \cite{bjork2002one}. We confirm the presence of the barriers using thermal activation measurements. Coupling the wires to evaporated aluminium electrodes, superconducting proximity effect is induced in the InAs segment, which we probe with our detector. We find substantial conductance suppression at subgap voltages originating from the proximitized wire segment. These barriers are important for monitoring subgap states both in qubits or Majorana devices.

\section{Experiments}

\subsection{Device outline}
\begin{figure}[h!]
	\includegraphics[width=\columnwidth]{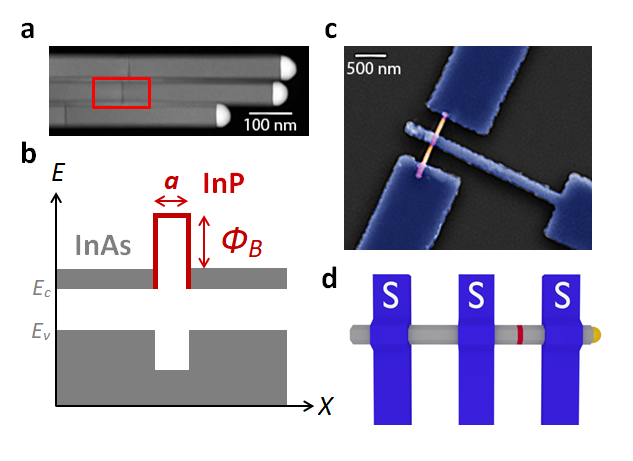}
	\caption{\label{fig:device}\textbf{Device structure}. \textbf{a} Transmission electron microscopy image of three InA nanowires deposited next to each other on a Silicon substrate. The dark regions correspond to the InP segments (see red rectangle). \textbf{b} Schematic view of the bandstructure of an InAs/InP/InAs heterostructure. The conduction band of the InAs is slightly filled ($E_F>E_C$), however due to the larger bandgap of InP it acts as a tunnel barrier with height of $\Phi_{\mathrm{B}}$ and width of $a$. A false colored SEM image of the device is shown in panel \textbf{c} and a simplified schematic view is given in panel \textbf{d}. The blue electrodes are superconducting aluminum (S). Two InAs segments are contacted by the electrodes, the right segment contains an InP barrier (red) and left one does not, which allows for comparative measurements.}
\end{figure}

We have used InAs nanowires with very narrow InP segments grown with Au assisted chemical beam epitaxy (CBE) \cite{zannier2019growth}. The nanowires had a diameter of apporoximately $50\,$nm, with a narrow  InP segment of thickness $a = 5.2\pm0.4\,$nm. Fig.~\ref{fig:device}a shows a trasmission electron microscopy (TEM) image of three InAs/InP nanowires where the InP segments are visible as a narrow dark region (also highlighted by the red rectangle). Previous studies using high resolution TEM images have shown, that the barriers are atomically sharp \cite{zannier2017nanoparticle,bjork2002one,bjork2003one,bleszynski2008imaging}.  Due to the large difference between the energy gaps of InAs and InP ($\sim1\,$eV), the InP segment will act as a sharp potential barrier for the electrons in the conduction band of the InAs nanowire as illustrated in the schematic view of the band structure of an InAs/InP/InAs heterostructure in Fig.~\ref{fig:device}b. 

To characterize the InP barriers, nanowires have been deposited on silicon substrate with a $290\,$nm thick oxide layer. Fig.~\ref{fig:device}c shows a false colored SEM image for a nanowire-based device with three superconductor aluminum electrodes (blue colored). Details of the fabrication are given in Methods.
The device has been fabricated such that the middle contact divides the nanowire into two segments: one of them contains the InP barrier, whereas the other section is a pure InAs segment for the purpose of comparative measurements (see Fig.~\ref{fig:device}d). 

\subsection{Thermal activation}
\begin{figure}[h!]
	\includegraphics[width=\columnwidth]{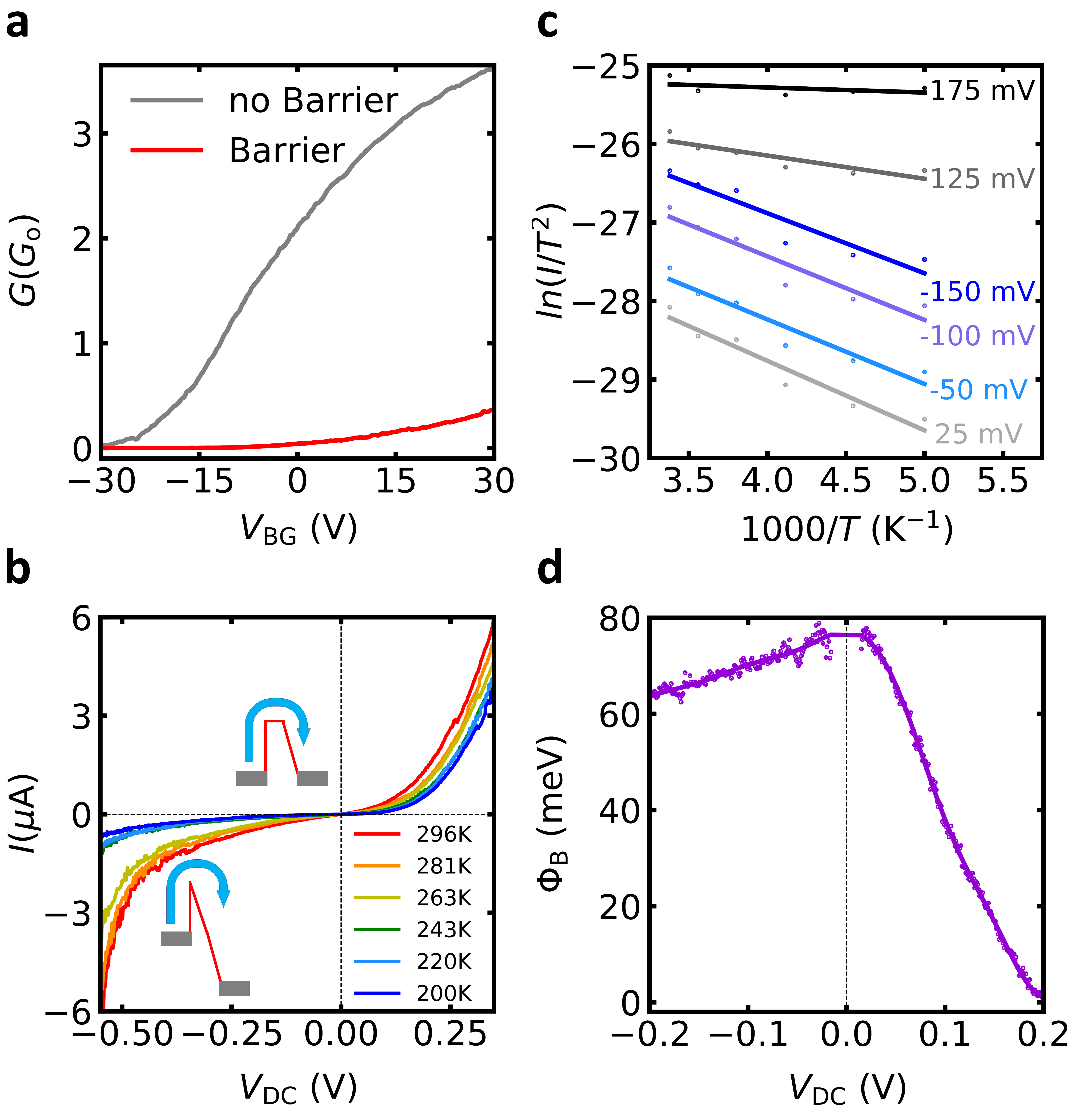}
	\caption{\label{fig:thermalactivation}\textbf{Characterization of tunnel barriers.} Panel \textbf{a} compares the conductance as a function backgate voltage for the segment with and without barrier at room temperature. The large decrease of conductance already at room temperature shows the presence of a barrier. Panel \textbf{b} shows $I-V_{DC}$ curves measured at different temperatures through the segment containing the barrier. The conductance at low dc bias is reduced and increases rapidly at high bias as a hallmark of tunnel barrier. In panel \textbf{c} the curves of panel \textbf{b} are replotted as $\mathrm{ln}(I/T^2)$ vs $\sim 1/T$ for different dc biases showing thermal activation behaviour. From the slope of the curves an effective barrier height can be extracted for different bias voltages, as shown in panel \textbf{d}. The effective barrier height at zero bias corresponds to the barrier height.}
\end{figure}

The backgate response for the two segments has been investigated by using standard lock-in technique. Fig.~\ref{fig:thermalactivation}a shows the measured conductance of the two segments as a function of backgate voltage at $T=300\,$K. Measurements at low temperature, $T=4\,$K  displaying similar features are shown in the supporting material. The gray line displays the conductance of the segment without the barrier, which shows a typical field-effect behaviour of InAs nanowires with a pinchoff voltage close to $V_{\mathrm{BG}}=-30\,$V. Compared to this, the segment hosting the InP barrier has substantially reduced conductance (see red curve). While the conductance of a fully open nanowire exceeds $3-6\,G_0(G_0=2e^2/h)$, the conductance of the barrier segment is limited to $0.4-0.5\,G_0$. At large gate voltages the nanowire segments contain already several highly transmitting modes, therefore the reduction in this case can be attributed to the presence of the barrier.
 
\noindent  To measure the effective height of this barrier, the $I-V$ characteristic measurements have been performed on the InP barrier segment at different temperatures as shown in Fig.~\ref{fig:thermalactivation}b. At higher temperatures electrons propagating in the InAs nanowire have sufficient energy to thermally go over the InP barrier, whereas at low temperatures the thermal activation is reduced. Using a simple model of a rectangular barrier the current $I$ through the barrier  as a function of temperature $T$ is given by: \cite{bjork2002one}
\begin{equation}
I=A_{\mathrm{r}}AT^2\exp{\left(-e\varphi_{\mathrm{B}}(V_{\mathrm{SD}})/k_{\mathrm{B}} T\right)},
\end{equation}
where $T$ is the temperature, $\varphi_{\mathrm{B}}(V_{\mathrm{SD}})$ is the DC bias dependent barrier height (see Fig.~\ref{fig:thermalactivation}b), $A$ is the nanowire cross-sectional area, $A_{\mathrm{r}}$ is Richardson constant and $k_B$ and $e$ are the Boltzmann constant and the electron charge, respectively. By plotting $\ln{I/T^2}$ vs $(1000/T)$ for the measured data as shown in Fig.~\ref{fig:thermalactivation}b for six different positive and negative bias voltages, the effective barrier height at given bias voltage can be estimated  directly from the slope of the fitted lines.  

The effective barrier height has been calculated for bias voltages in the window of $-0.2$ to $0.2\,$V as shown in Fig.~\ref{fig:thermalactivation}d. A narrow window of $\pm 15$ mV around zero bias has been excluded from the analysis due to the small values of the current leading to large scattering on the effective barrier height. The estimated effective barrier height at zero bias is around $80\,$meV. One can also notice an asymmetry in the effective barrier height between positive and negative bias voltages. The lack of symmetry in bias voltage is already present in the $I-V$ curves of panel b, where the breakdown voltage is substantially different for forward and reverse bias. A possible reason for this might come from a structural asymmetry\cite{nylund2016designed} : during the growth process the transition from InAs to InP is sharper than for the transition InP to InAs as shown by high resolution TEM in Ref.~\citenum{zannier2017nanoparticle} and illustrated by the insets of panel b.

A second thing to notice is that the effective barrier height is smaller than it is expected from workfunction arguments and also substantially smaller than measured in Ref. \citenum{bjork2002one,bjork2003one} with InP  barrier thickness about 80 nm. The reason for this small value might come from the oversimplification of our model. First, the model use a rectangular shaped barrier, which might not be realistic for such thin barriers, where the  transition from InAs to InP is atomically sharp. The other one (InP-to-InAs) is slightly graded, and the transition is on a non-negligible length scale (about 5 nm). Moreover, the model neglects changes in the barrier shape due to the applied voltage voltage and also neglects the effect of quantum tunneling, which both are relevant for such  narrow barriers contrary to wide barriers studied before in Refs.~\citenum{bjork2002one,bjork2003one,nylund2016designed}.

\subsection{Low temperature superconducting spectroscopy}
\begin{figure}[tb!]
	\includegraphics[width=\columnwidth]{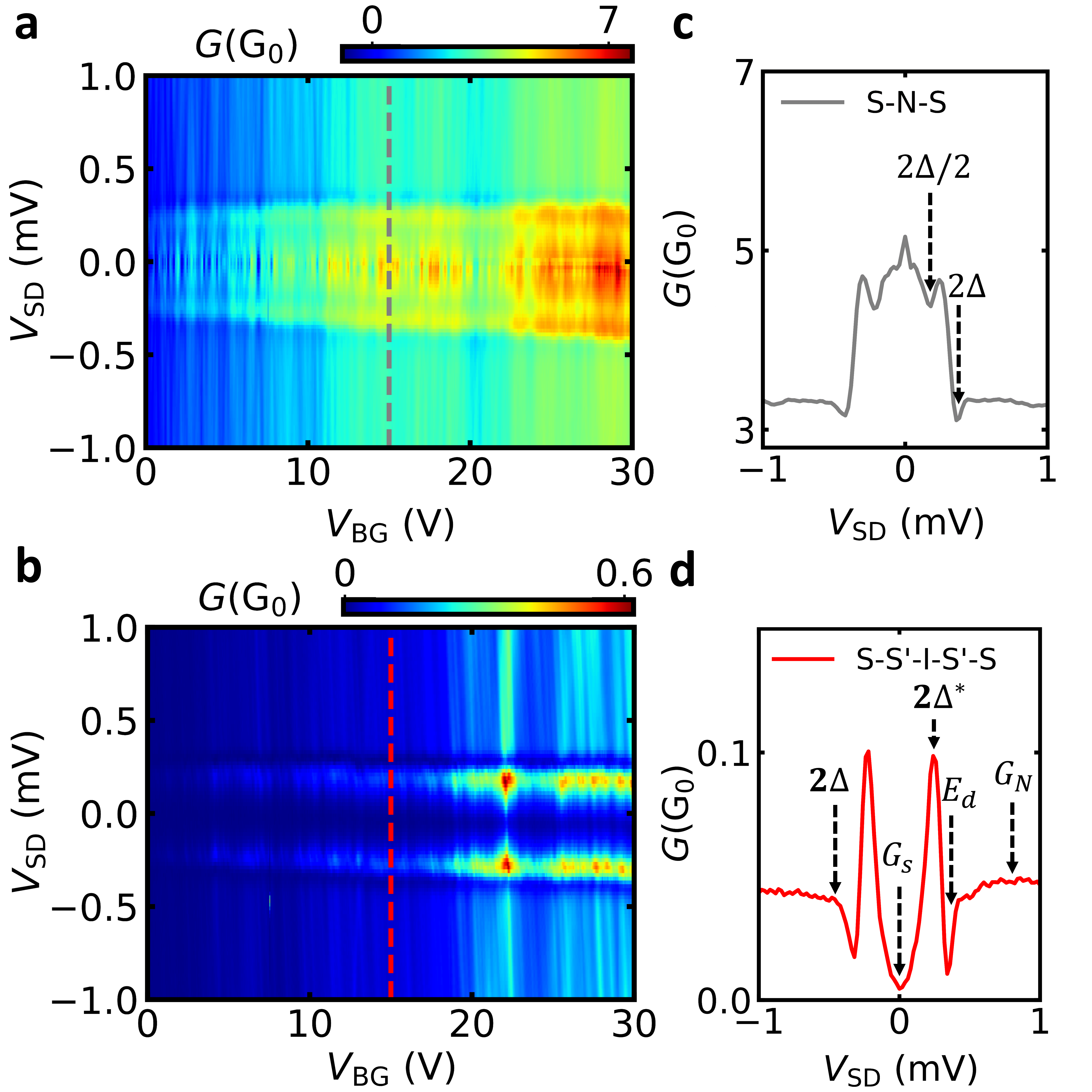}
	\caption{\label{fig:2dmaps}\textbf{Low temperature superconductivity measurements.} Differential conductance as a function of $V_{\mathrm{SD}}$ and $V_{\mathrm{BG}}$ for InAs segment without barrier and with barrier shown in panel \textbf{a} and \textbf{b}, respectively. Corresponding cuts from panel \textbf{a}  and panel \textbf{b} at the positions of the dashed lines are shown on panel \textbf{c} and \textbf{d}, respectively. The structure without barrier on panel \textbf{c} shows a clear enhancement at low bias voltages and the presence of multiple Andreev reflection marked by arrows on panel \textbf{c}. The peak at zero bias results from supercurrent between the two electrodes. On the contrary panel, \textbf{d} shows a presence of large conductance peaks at voltages marked as $2\Delta^{*}$ and a strong decrease of conductance within the gap. Unexpectedly, small dips next to the peaks appear marked with the $E_{\mathrm{d}}$. The figure also marks the values of sub-gap and normal state conductance ($G_{\mathrm{S}}$ and $G_{\mathrm{N}}$) used in further analysis.   }
\end{figure}

Superconducting spectroscopy using built-in barriers were performed in dilution refrigerator at temperature of 30 mK. A small, 10 $\mu$V AC signal  has been applied to the middle contact and the differential conductance in both segments has been measured as a function of the backgate voltage $V_{\mathrm{BG}}$ and DC bias $V_{\mathrm{SD}}$. Fig.~\ref{fig:2dmaps}a and b show measurements on device segments without barrier and with barrier, respectively, and corresponding line cuts at $V_{\mathrm{BG}}=15\,$V are presented in Fig.~\ref{fig:2dmaps}c,d respectively. For the two segments the conductance behaves differently in the subgap region($|V_{\mathrm{SD}}|<0.3\,$meV).
In the absence of the barrier the conductance is enhanced in the subgap regions as expected for well transmitting \emph{SNS} junction\cite{jespersen2009mesoscopic}. On the other hand, the segment containing the barrier shows a large suppression at the same bias window due to the low transmission probability for electrons through the InP barrier.\par

\noindent The measurements of the segment without barrier show existence of multiple Andreev reflections (MAR) between the two superconducting contacts indicated by the presence of  dips in the sub-gap conductance\cite{gunel2014crossover} at $\pm2\Delta/e=\pm400\,\mu$V, $\pm2\Delta/2e$ marked by arrows in Fig.~\ref{fig:2dmaps}c,  where $\Delta$ is the gap of the superconducting electrode. Measurements of another junction displaying also MAR features at $\pm2\Delta/3e$ are presented in the supporting information. 

On the contrary, the characteristics of the segment with InP barrier is different (see Fig.~\ref{fig:2dmaps}d), it has a superconducting gap-like feature: large peaks reminiscent of the superconducting coherence peaks and reduced conductance in-between. However, compared to the DOS of metallic superconductors also differences can be seen: i) The peaks appear at reduced energy marked with $2\Delta^{*}=~\pm220\,\mu$eV. ii) At energies above the gap a small dip appears, marked by $E_d=~\pm330\,\mu$eV. iii) The sub-gap conductance remains finite even at zero bias voltage. This \emph{softness} of the gap can be characterized by the sub-gap to normal state conductance ratio\cite{chang2015hard}, $G_{\mathrm{S}}/G_{\mathrm{N}}$ which is $\sim 0.1$ in our case, where $G_{\mathrm{S}}=G(V_{\mathrm{SD}}=0)$ and $G_{\mathrm{N}}$ is the normal state conductance averaged in a small bias window above the gap (see Fig.~\ref{fig:2dmaps}d).

\section{Discussion}

In the following we will concentrate on the analysis of the low-temperature spectroscopy measurements. It is expected that when a superconducting contact is placed on an InAs wire an induced proximity superconducting region appears in the wire in the vicinity of the electrode. Therefore our structure containing the barrier can be modelled as \emph{S-$S'$-I-$S'$-S}, where $I$ marks the barrier, and $S'$ marks the induced superconducting region in the wire. It is known that in the long junction limit ($L>\xi$, where $L$ is the junction length and $\xi$ is the superconducting coherence length) in $S'$ the superconducting gap becomes reduced as the distance from the SC interface of the superconductor is increased\cite{zhitlukhina2016anomalous}. In our structure this happens on both sides of the barrier. Therefore our structure to first order can be understood as one proximity induced superconducting region probes the other. The correctness of this picture can be verified with devices, where one of the superconducting electrode is replaced by the normal one, hence an \emph{S-$S'$-I-N} structure is formed. Measurements on such a structure are given in the supporting material and show similar features, but at smaller energy scale (since only one superconducting electrode is present), and with reduced energy resolution.

\begin{figure}[h!]
	\includegraphics[width=\columnwidth]{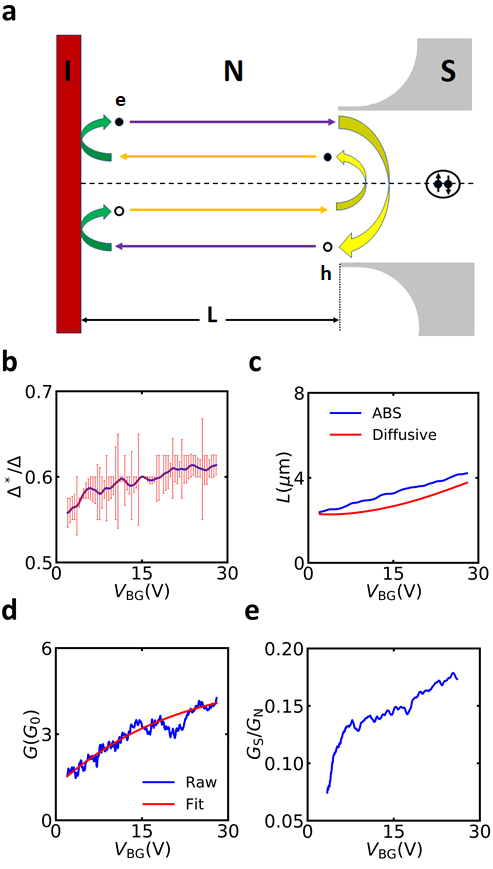}
	\caption{\label{fig:analysis}\textbf{Induced superconductivity}  Panel \textbf{a} shows schematics of the formation of Andreev Bound states between an the superconducting contact and the tunnel barrier. Empty circles represent holes, whereas filled circles represent electronic states. The loop corresponds to the lowest order process to form Andreev Bound states. The phases accumulated during the round trip are given by Eq.~\ref{Eq.Phase}.   Panel \textbf{b} shows the extracted superconducting gap normalized by the bulk gap of aluminum. For $\Delta$ we have used $200\,\mu$V. Panel \textbf{c} shows $L_{\mathrm{tr}}$ (in blue) extracted from the Andreev Bound state energy using Eq.~\ref{Eq.Phase} as a function of backgate voltage. The red curve shows the path length than an electron travels with diffusion coefficient $D$ from the superconducting contact to the barrier as a function of backgate voltage. Panel \textbf{d} shows the normal state conductance extracted as from Fig.~\ref{fig:2dmaps}a at $V_{\mathrm{SD}}=1\,$mV. Panel \textbf{e} shows the subgap conductance normalized by the normal state conductance as defined in Fig.~\ref{fig:2dmaps}e.  }
\end{figure}

In few channel systems the proximity effect is often explained in the framework of Andreev Bound States (ABS)\cite{pannetier2000andreev, Klapwijk2004}. In our geometry, ABSs can be formed between the superconducting electrode and the barrier. The formation of the ABS is shown in Fig.~\ref{fig:analysis}a. An electron travelling in the wire is normal reflected from the barrier, whereas it is Andreev reflected from the superconductor. The same holds for a hole. After two normal and two Andreev reflections a bound state can be formed. Using the Bohr-Sommerfeld quantization the phase factor acquired by the wave function during the full cycle should be multiple of $2\pi$:

\begin{equation}
    \label{Eq.Phase}
    2(k_{\mathrm{e}}-k_{\mathrm{h}})L-2\arccos\left(\frac{E}{|\Delta|}\right)=2\pi n,
\end{equation}
where $k_{\mathrm{e}}$, $k_{\mathrm{h}}$ are the wavenumbers of the electron and the hole, $L$ is the distance between the SC electrode and and the barrier, $E$ is the bound state energy, $\Delta$ is the superconducting gap and $n$ is an integer number. In the short junction limit the first term can be omitted, resulting in $E= \Delta$, however, here this is not the case. $\Delta k=k_{\mathrm{e}}-k_{\mathrm{h}}$ can be approximated from the dispersion relation: $\Delta k = E/(\hbar v_{\mathrm{F}})$, where $v_{\mathrm{F}}$ is the Fermi velocity.
Similar equations can be written up for the other side of the barrier. Here for simplicity it is assumed that the energy of the ABS is the same on both sides and within this picture we did not consider bound states connecting the two sides.

In the framework of ABS the coherence peaks at $V_{\mathrm{SD}}=2\Delta^* /e$ in Fig.~\eqref{fig:2dmaps}d  appear, when the applied bias voltage aligns the occupied ABSs (at energy $-E$) on one side of the barrier with the unoccupied ABS (at $+E$) on the other side of the barrier. Thus the coherence peak position is directly related to the energy of the ABS, i.e. $\Delta^*=E$. Fig.~\ref{fig:analysis}b shows the extracted $\Delta^*$ normalized with the bulk gap as a function of the back gate voltage. A small, but steady increase is shown as a function of the gate voltage. Let us estimate now, the trajectory length based on Eq.~\ref{Eq.Phase}. The control segment without the barrier can be used to extract the density and Fermi velocity as a function of back-gate voltage. The conductance of the segment without the barrier, which is used for obtaining these parameters is shown on panel d of Fig.~\ref{fig:analysis}. This was measured at a DC bias voltage of 1$\,$mV to avoid superconducting features. Using the ABS energy and the Fermi velocity extracted from the reference junction (given in the Supporting information, with typical values $v_{\mathrm{F}}=10^6\,$m$/$s) and  the  length of the junction can be extracted using Eq.~\eqref{Eq.Phase} with n=0. This is shown in Fig.~\ref{fig:analysis}c as a function of backgate voltage. The extracted trajectory length is $3-4\,\mu$m, which is  much larger than the nominal length of the InAs segment between the barrier and S electrode, which is approximately $150\,$nm. The reason for this discrepancy lies in the assumption that our junction is ballistic. Again, we can use the control junction to extract a mean-free path (see supporting information) which is in the order of $20-30\,$nm.  We can also extract the diffusion coefficient, $D$ from the control junction, using Einstein relation and a 3D density of states for the nanowire (shown in supporting information), which results in values in the range of $D=0.008-0.011\,$m$^2/$s. With the diffusion constant the effective trajectory length an electron undertakes during its diffusive motion from the SC electrode to the barrier can be calculated: $L_{\mathrm{tr}}=\frac{L^2}{D} v_{\mathrm{F}}.$

This $L_{\mathrm{tr}}$ is plotted in Fig.~\ref{fig:analysis}c with red and is in quite good agreement with the blue trace extracted from the ABS energy. This good agreement points to the direction that although ABSs give a conceptually a simple description of the induced gap, several scattering events take place as an electron propagate from the barrier into the S electrode. This could originate from finite reflection at the S - nanowire interface or from diffusive transport in the nanowire. 
Note, in our nanowires there are several conductance channels leading to various ABS trajectories, which can be described by a distribution of energies of ABSs. In this case $L$ derived from $\Delta^*$ captures the most probable trajectory length.  Thus assuming only a single channel and using Eq.~\eqref{Eq.Phase} is a valid simplification to get an estimate for the typical trajectory length.

The ABS energy, or with other words the induced gap, shows an increase as a function of the backgate voltage (see Fig.~\ref{fig:analysis}b). Similar tendency was observed in Ref.~\citenum{Juenger2019}, where an increase at low and a clear saturation at larger backgate voltages was seen. In this study they attributed this effect to the transition from long ballistic to the short junction regime, where the transition was driven by the Fermi velocity change induced by the density change. In our case, the ABS energies are increasing as a function of gate voltage in the full range investigated. The junction is in the  transition region between long diffusive and short regime ($L\approx \xi$, see Supporting info), but no clear signatures of this transition was observed. However, a strong increase and saturation behaviour was seen for another, \emph{S-$S'$-I-N} junction shown in the supporting information, where this behaviour might be attributed to the transition from long diffusive to short junction regime.

Finally, we comment on the gap suppression. The value of subgap conductance normalized by the normal state conductance is shown in Fig.~\ref{fig:analysis}e for the entire gate range. This shows an increase as a function of backgate voltage from $0.08$ to $0.17$. The increase can be understood as a reduction of barrier height as the electron density increases.  These low values are quite promising for future experiments. This sub-gap suppression is much larger than what was found in Ref.~\citenum{Juenger2019} where the barrier was defined by the change of crystallographic structure of the wire and is larger than what one can usually achieve using gate defined tunnel barriers.

Whereas the values are encouraging for future experiments, for high spectral resolution an even larger sub-gap conductance suppression would be desirable. It is expected that if the normal state conductance of the barrier is proportional to $\mathcal{T}$, the transmission value, then the sub-gap conductance scales with $\mathcal{T}^2$, since two electron charges are transmitted for Andreev reflection based transport processes. Therefore by increasing barrier width (the width of the InP segment) a smaller $\mathcal{T}$, and therefore larger conductance suppression is expected. Also, since the superconducting contact placed on the wire has finite $150-200\,$nm extension this results in electrons injected at different distances from the barrier. This results in electron trajectories with different length, and lead also to different ABS energies, hence to a smearing of the gap and reducing subgap suppression. Finally, subgap states present in InAs nanowires discussed in Ref.\citenum{Takei2013} e.g. stemming from interface inhomegenity can also give contributions to the sub-gap conductance. However, their role could be better assessed using barriers with smaller transmission.

So far we could describe most of the features in Fig.~\ref{fig:2dmaps}d based on Andreev bound states with an effective trajectory length.  In Ref.~\citenum{zhitlukhina2016anomalous} detailed calculations for similar structures have been made based on scattering formalism. These calculations match well with our measurements, even reproducing the dip above the gap (at $E_{\mathrm{d}}$). This dip usually appears when localized states, resonances are present between the barrier and the superconducting electrodes, in \emph{SIS} or \emph{SIN} structures\cite{zhitlukhina2016anomalous}. In our case the ABS can play the role of the resonance, which is underlined by the aformentioned detailed theoretical calculation.

\section{Conclusions}

In summary, we have shown that InP segments embedded in InAs nanowires can be used as tunnel probes for probing SC bound states in the proximitized wire regions. We have used thermal activation measurements to confirm the presence of the InP barriers. Low temperature transport measurements have shown an induced gap-like structure which we have described with the presence of Andreev Bound states in the lead segment. We have seen that the induced gap is gate tunable, originating from the change $D$ (or Thouless energy). The curves showed a sub-gap conductance suppression below $0.1$, which is promising for future applications. Compared to quantum dot based probes, or probes based on crystal phase engineering our InP barrier does not contain internal resonances, therefore will not hybridize with the states to be probed. These InP segments  offer sharp tunnel barriers, where the transmission can be changed by changing the width of the barrier during the growth. We have used InP segments with thickness of $5.2\,$nm, which lead to a subgap suppression of $0.1$ (see Supporting Information). We have estimated the transmission of the barrier is $0.02-0.1$. Further increase of the subgap suppression could be achieved by increasing the barrier width. However, in the literature the presence of a soft gap has been discussed. This soft gap might also originate from sub-gap states present in the proximity induced region, not from the imperfection of the tunnel probe. The presence of such sub-gap states are important both for qubit or for Majorana devices. With further increase of the tunnel barrier the presence of these states could be studied. Moreover, it has been claimed, that using in-situ grown Al contacts a better controlled proximity region could be achieved, with a reduced number of sub-gap states\cite{chang2015hard, Krogstrup2015}. Therefore, it would be beneficial to combine these in-situ grown wires with Al shell with the InP tunnel barriers which would also allow better probing of exotic bound states.

\section*{Methods}

InAs/InP nanowires heterostructure of diameter approximately $50\,$nm with a very narrow  InP segment of thickness about $a = 5.2\pm0.4\,$nm have been grown by Au assisted chemical beam epitaxy (CBE) on InAs(111)B substrates, using trimethylindium (TMIn), tert-butylarsine (TBAs) and tert-butylphosphine (TBP) as metalorganic precursors, with line pressures of 0.2, 0.8 and 4 Torr, respectively\cite{zannier2019growth}. Then nanowires have been deposited on silicon substrate  with a $290\,$nm oxide layer by means of mechanical manipulator. The device has been fabricated by electron beam lithography (EBL) then Ti/ Al layers have been deposited with thicknesses $10$/$80\,$nm. Prior to the deposition, the oxide formed on the surface of the InAs has been removed with Ar bombardment. For the \emph{SIN} structures a separate step e-beam step has been performed to realize the Ti/Au contacts (10/80$\,$nm). The barriers have been located using SEM images, and the growth parameters (distance from gold catalyst).

Low temperature measurements have been done in a Leiden Cryogenics CF-400 top loading cryo-free dilution refrigerator system with a base temperature of $30\,$mK. We have used standard lock-in technique at 137 Hz by applying very small AC signal with 10 $\mu$V amplitude on the middle contact and measuring the conductance of the two segments via home-built I/V converters as a function of DC bias voltage and backgate voltage.

\section*{Author contributions}
T.E. and O.K., L.I., P.S. fabricated the devices, T.E. and O.K. and Z. S. performed the measurements and did the data analysis. L. S., F. R. and V. Z. grew the wires. All authors discussed the results and worked on the manuscript. Sz. Cs. and P.M. guided the project. The data in this publication are available in numerical form at: \url{https://doi.org/10.5281/zenodo.3519430}  

\section*{Acknowledgments}
We acknowledge C. J\"unger, A. Baumgartner, A. Palyi, J. Nygard, A. Virosztek, T. Feher for useful discussions and M. G. Beckerne, F. Fulop, M. Hajdu for their technical support.
This work has received funding Topograph FlagERA , the SuperTop QuantERA network, the FET Open AndQC and from the OTKA FK-123894 grants. P.M. acknowledges support from the Bolyai Fellowship, the Marie Curie grant and we acknowledge the National Research, Development and Innovation Fund of Hungary within the Quantum Technology National Excellence Program (Project Nr. 2017-1.2.1-NKP-2017-00001).

\nocite{*}
\bibliography{main}

\end{document}



\title{Supporting info for probing proximity induced superconductivity in InAs nanowire using built-in barriers}


\author{Tosson Elalaily}
\affiliation{Department of Physics, Budapest University of Technology and Economics and Nanoelectronics 'Momentum' Research Group of the Hungarian Academy of Sciences, Budafoki ut 8, 1111 Budapest, Hungary}
\affiliation{Department of Physics , Faculty of Science, Tanta University, Al-Geish St., 31111 Tanta, Gharbia, Egypt.}

\author{Oliv\'er K\"urt\"ossy}
\affiliation{Department of Physics, Budapest University of Technology and Economics and Nanoelectronics 'Momentum' Research Group of the Hungarian Academy of Sciences, Budafoki ut 8, 1111 Budapest, Hungary}

\author{Valentina Zannier}
\affiliation{NEST, Istituto Nanoscienze-CNR and Scuola Normale Superiore, Piazza San Silvestro 12, I-56127 Pisa, Italy}

\author{Zolt\'an Scher\"ubl}
\affiliation{Department of Physics, Budapest University of Technology and Economics and Nanoelectronics 'Momentum' Research Group of the Hungarian Academy of Sciences, Budafoki ut 8, 1111 Budapest, Hungary}

\author{Istv\'an Endre Luk\'acs}
\affiliation{Center for Energy Research, Institute of Technical Physics and Material Science, Konkoly-Thege Mikl\'os \'ut 29-33., H-1121, Budapest, Hungary}

\author{Pawan Srivastava}
\affiliation{Department of Physics, Budapest University of Technology and Economics and Nanoelectronics 'Momentum' Research Group of the Hungarian Academy of Sciences, Budafoki ut 8, 1111 Budapest, Hungary}

\author{Francesca Rossi}
\affiliation{IMEM-CNR, Parco Area delle Scienze 37/A, I-43124 Parma, Italy}

\author{Lucia Sorba}
\affiliation{NEST, Istituto Nanoscienze-CNR and Scuola Normale Superiore, Piazza San Silvestro 12, I-56127 Pisa, Italy}

\author{Szabolcs Csonka}
\email{szabolcs.csonka@mono.eik.bme.hu}
\affiliation{Department of Physics, Budapest University of Technology and Economics and Nanoelectronics 'Momentum' Research Group of the Hungarian Academy of Sciences, Budafoki ut 8, 1111 Budapest, Hungary}

\author{P\'eter Makk}
\email{peter.makk@mail.bme.hu}
\affiliation{Department of Physics, Budapest University of Technology and Economics and Nanoelectronics 'Momentum' Research Group of the Hungarian Academy of Sciences, Budafoki ut 8, 1111 Budapest, Hungary}

\date{\today}

\maketitle

\section{4K conductance comparison}
\begin{figure}[tb]
	\includegraphics[width=.4\columnwidth]{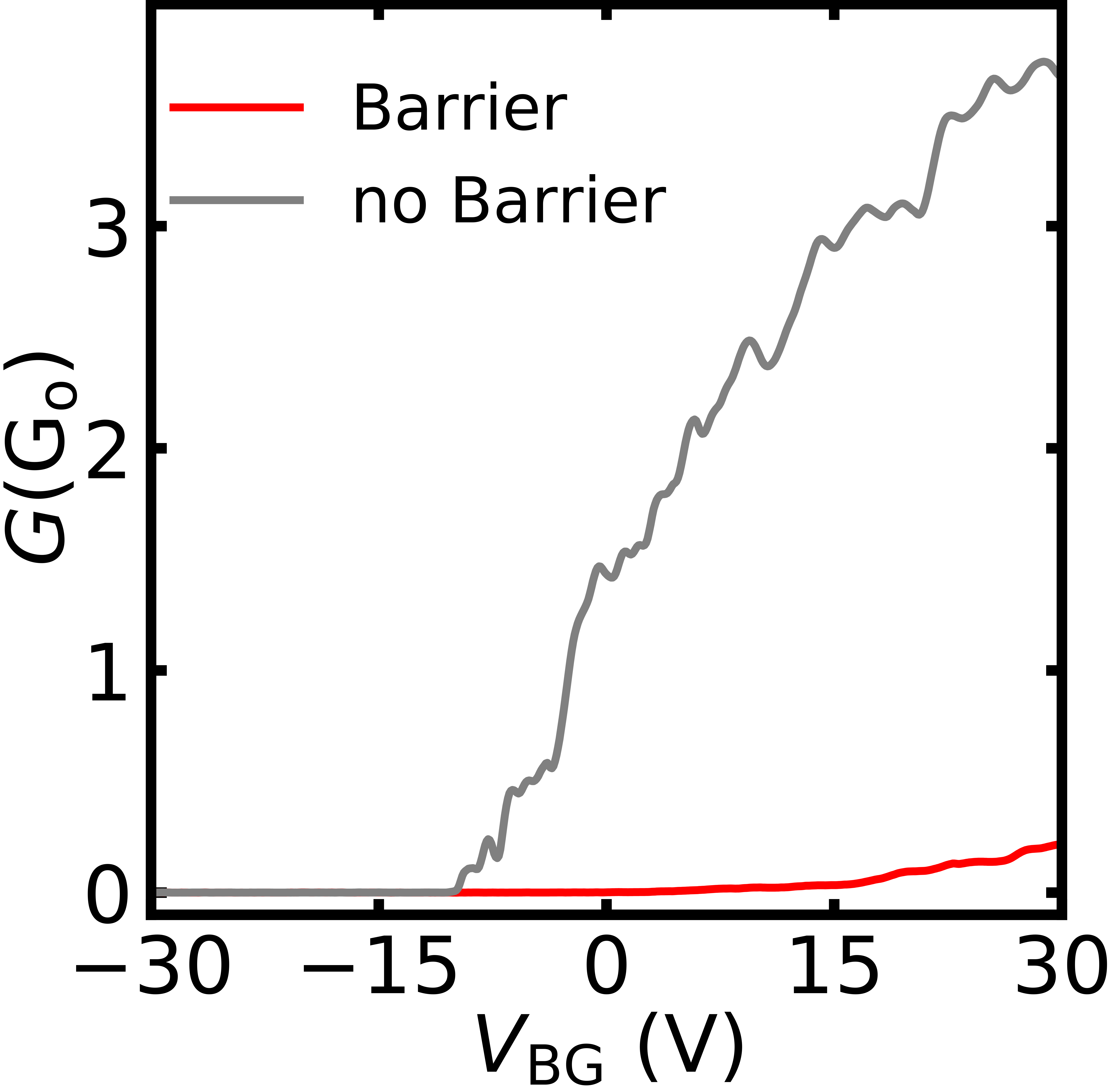}
	\caption{\label{fig:Trace4K_Supporting}\textbf{}  Conductance vs $V_{\mathrm{BG}}$ for the same device as in the main text measured at $T=4\,$K. The segments with and without barrier show a clear difference. }
\end{figure}

Here, in Fig.~\ref{fig:Trace4K_Supporting} the $4\,$K conductance as a function of backgate voltage is shown for the nanowire presented in the main text.
The gate traces show a similar trend to the gate traces measured at room temperature: there is a large difference between the conductance of the two segments due to the presence of the barrier in one of the two InAs segments.  

Based on similar measurements of 7 nanowires, 
the conductance of wire segments without barrier sections has at $V_{\mathrm{BG}}=30\,$V a mean value and standard deviation of $2.7\,G_0$ and $0.71\,G_0$ respectively, while for the segment with barrier it $0.289\,G_0$ and $0.245\,G_0$ respectively.  The scattering of the barrier transmission is consistent with the variation of width of the InP barrier,  which changes from wire to wire in the range of $5.2\pm 0.4\,$nm according to TEM measurements.

The tunnel barrier reduces the conductance with more than an order of magnitude, generating a bottleneck in the  transport. Thus the conductance sensitively measures the transport signature of the tunneling processes.

At $T=4\,$K  the thermal activation is suppressed, thus the transport through the barrier is dominated by quantum tunneling. Based on the typical tunnel conductance of 0.2\,$G_0$, we could estimate the transmission of the tunnel barrier. When the nanowire is opened by large gate voltage $V_{\mathrm{BG}}\geq 30\,$V, we expect 5-10 conductance channels in the InAs segment. Since the InP barrier is sharp, we expect that all channels contribute to the tunneling. Therefore the transmission for a single channel has a typical value of $0.02-0.03$ with an upper limit of $0.1$.  This low transmission strongly suppresses  Cooper pair tunneling, and also isolates bound states on the two sides of the barrier.

\section{Second device without barrier - MAR}

\begin{figure}[tb]
	\includegraphics[width=.8\columnwidth]{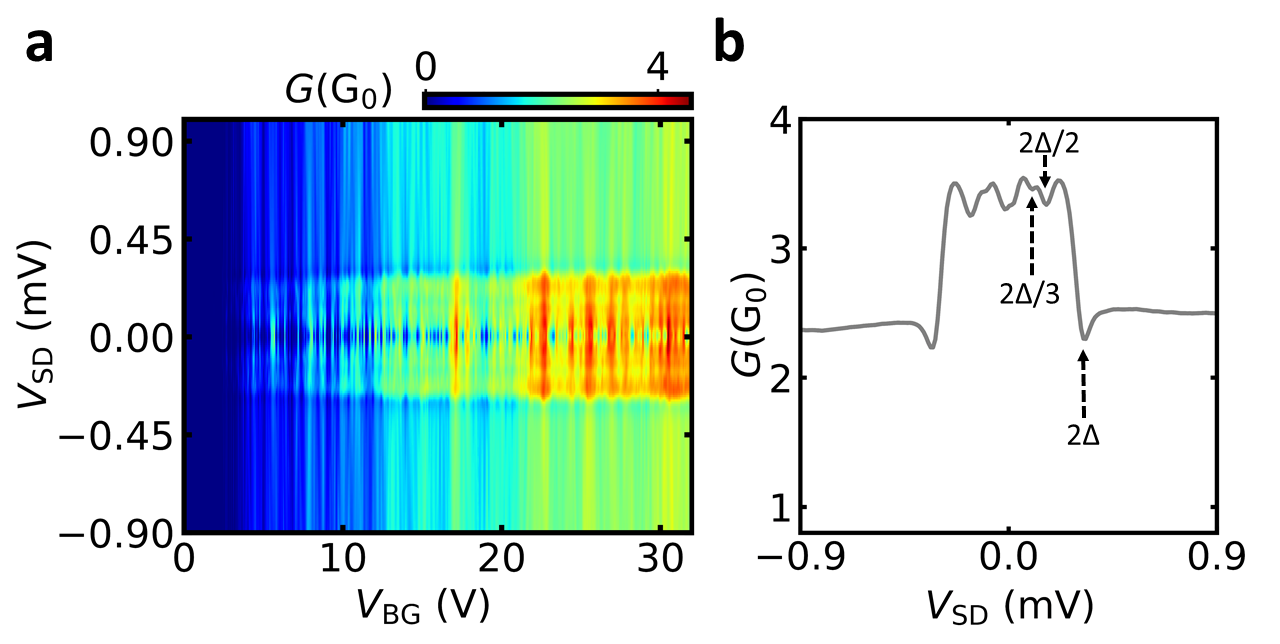}
	\caption{\label{fig:MAR_Supporting}\textbf{}  \textbf{(a)} Differential conductance as a function of $V_{\mathrm{SD}}$ and $V_{\mathrm{BG}}$ for another nanowire segment without barrier with two S contacts. \textbf{(b)} The line cut at $V_{\mathrm{BG}}=32\,$V shows clearly the presence of multiple Andreev reflections 
	in the sub-gab conductance at positions $\pm2\Delta/e=~\pm330\,\mu$V, $\pm2\Delta/2e$ and $\pm2\Delta/3e$ marked by arrows. An averaging over a gate range of 0.2 V has been performed to suppresses conductance fluctuations.}
\end{figure}

Fig.~\ref{fig:MAR_Supporting} shows a low-temperature measurement for a second device without InP barrier. Panel (a) shows the conductance as a function of bias voltage and backgate, showing a large conductance enhancement at low bias. Panel (b) shows a linecut for fixed gate voltage $V_{\mathrm{BG}}=32\,$V. Clear signs of Multiple Andreev reflections are present at positions $\pm2\Delta/e=~\pm330\,\mu$V, $\pm2\Delta/2e$ and $\pm2\Delta/3e$.

\section{N-I-S measurements}

\begin{figure}[tb]
	\includegraphics[width=.8\columnwidth]{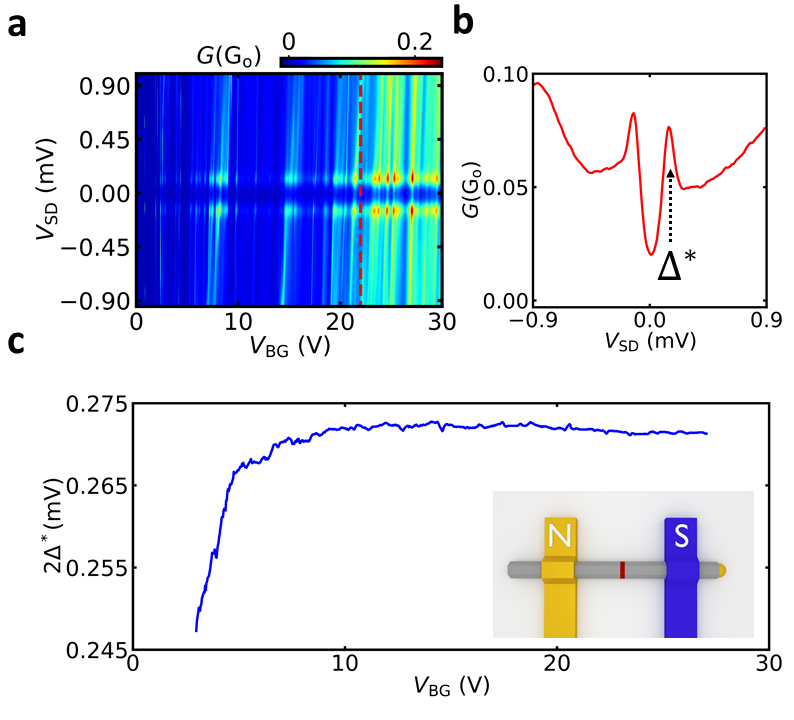}
	\caption{\label{fig:NIS_Supporting}\textbf{} (a) Differential conductance as a function of $V_{\mathrm{SD}}$ and $V_{\mathrm{BG}}$ for segment with the InP barrier for other nanowire device with \emph{NIS} configuration. \textbf{(b)} Line cut at $V_{\mathrm{BG}}=22\,$V shows the decrease in the sub-gab conductance due to the presence of the barrier. \textbf{(c)} The energy of the  peak position marked in panel b) as a function of backgate voltage. Inset: a schematics of the measured device.}
\end{figure}

To check the coupling between proximity regions formed on the opposite sides of the tunnel barriers we have fabricated devices by replacing one of S contacts with normal one (Ti/Au 10/80nm) , thus forming \emph{S-$S'$-I-N} structures. The inset of Fig.~\ref{fig:NIS_Supporting}c shows the concept of the device under consideration. The differential conductance as a function of $V_{\mathrm{SD}}$ and $V_{\mathrm{BG}}$ is presented in Fig.~\ref{fig:NIS_Supporting} measured at $T=30\,$mK. Comparing the subgap features for \emph{S-$S'$-I-N} (Fig.~\ref{fig:NIS_Supporting}a and b)  with the subgap characteristic of \emph{S-$S'$-I-$S'$-S} (see Fig.~3b and d in Main Text),  we observed very similar features. Due to the presence of the barrier the sub-gap conductance is decreased  and  a coherence peak due to the ABS is observed at $\Delta^*$. The structures are sharper for  \emph{S-$S'$-I-$S'$-S}, since the superconducting DOS provides better energy filtering. Panel (c) shows conductance peak  position, $2\Delta^*$ extracted from panel (a).  The value of $\Delta^*$ as the backgate voltage increases is followed by a saturation at large gate voltage. A similar trend was observed in Ref.~\onlinecite{Juenger2019}, where they attributed this to a transition from a long ballistic to a short junction regime. Similar transition might happen here, however, for our devices this transition takes place rather from a long diffusive to the short regime (see also discussion in the next section).

\section{Density and Fermi velocity calculation for the reference segment}

In this section we describe the capacitive model used in the main text and also the diffusive model used to extract $L_{\mathrm{tr}}$.

\begin{figure}[tb]
	\includegraphics[width=.8\columnwidth]{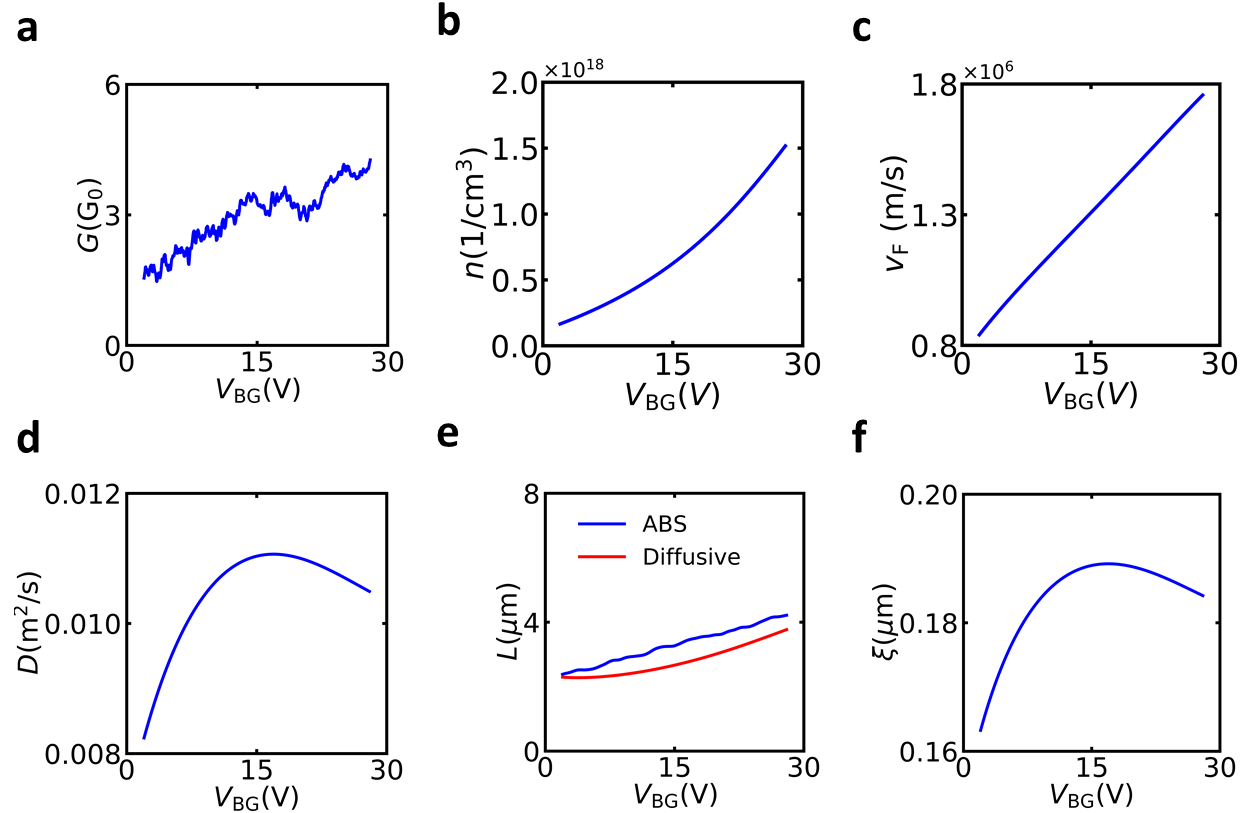}
	\caption{\label{fig:Fig4_supporting_2}\textbf{} (a) Conductance as a function of back gate voltage at $T=30\,$mK and $V_{\mathrm{SD}}$ =1 mV (the same as in Fig.4d of the main text).  \textbf{(b)} Electron density, \textbf{(c)} Fermi velocity \textbf{(d)} and diffusion constant \textbf{(e)} calculated using the capacitance model explained in the text. Panel \textbf{(e)} shows  $L_{\mathrm{eff}}$ in blue extracted from the Andreev Bound state energy using Equation~\ref{eq:abs_condition_ordered_effective} as a function of backgate voltage. The red curve shows the path length obtained from the diffusive model using Eq.~\ref{eq:path}. \textbf{(f)} Diffusive coherence length as a function of the backgate voltage is shown.}
\end{figure}

To extract the electron density we measured the conductance of the reference segment without the InP barrier. This segment had the same geometry as the device with the barrier  (see Fig. 1c and d in the main text).  The conductance of the reference segment as function of back gate voltage is shown on Fig.~\ref{fig:Fig4_supporting_2}a.  We used the Drude model to extract the electron density: 

\begin{equation}\label{eq:drude}
n=\frac{\sigma}{e\mu},
\end{equation}
\noindent where $\mu$ is the electron mobility, $\sigma$ is the conductivity. The conductivity is given by

\begin{equation}\label{eq:sigma}
\sigma=\frac{2LG}{R^2\pi},
\end{equation}
where $R$ is the wire diameter, $2L$ is the segment length (twice the distance $L$ between superconductor contact and the barrier in the segment with barrier as mentioned in the main text), and $G$ is the measured conductance. The mobility can be expressed via the transconductance following Ref.~\onlinecite{Schroer2010},

\begin{equation}\label{eq:mu}
\mu=\frac{\partial G}{\partial V_\mathrm{{BG}}}\frac{4L^2}{C_\mathrm{{BG}}},
\end{equation}
where $\frac{\partial G}{\partial V_\mathrm{{BG}}}$ is the transconductance calculated numerically from the conductance as a function of gate voltage. $C_{\mathrm{BG}}$ is the capacitance beteen the backgate and nanowire, which can be estimated with the capacitance of an infinitely long wire inside a dielectric parallel to a plane (Ref.~\onlinecite{Wunnicke2006}):

\begin{equation}\label{eq:c}
C_\mathrm{{BG}}=\frac{4\epsilon_\mathrm{0}\epsilon_\mathrm{r}L}{\mathrm{acosh(\frac{t+R}{R})}}.
\end{equation}
Here $t$  is the distance between the wire and the the backgate plane, $\epsilon_{\mathrm{r}}$ is the dielectric constant of the back gate. In our case $t= 290$ nm and $\epsilon_{\mathrm{r}}= 3.9$. For the calculation the conductance trace of panel (a) was fitted with a 3rd order polynomial. The density calculated using Equations~1-4 is plotted on Fig.~\ref{fig:Fig4_supporting_2}b.

Assuming isotropic dispersion, the charge density is proportional to the Fermi wavenumber:

\begin{equation}\label{eq:kf}
n=\frac{k_\mathrm{F}^3}{3\pi^2}.
\end{equation}
Using a parabolic dispersion relation

\begin{equation}\label{eq:vf}
v_\mathrm{F}=\frac{\hbar k_\mathrm{F}}{m^*}=\frac{\hbar}{m^*}(3n\pi^2)^{\frac{1}{3}},
\end{equation}
where $m^*$ is the effective mass, $m^*=0.023m_e$. 
The Fermi velocity as a function of backgate voltage is shown in Fig.~\ref{fig:Fig4_supporting_2}c. This was used in the main text to extract the effective length of the segment with barrier using: 

\begin{equation}\label{eq:abs_condition_ordered_effective}
\frac{E}{\vert \Delta \vert}=\cos\left( \frac{2L}{\hbar v_\mathrm{F}}E-\pi n\right) ; \ \ \ n\in \mathbb{Z}.
\end{equation}
The result is shown again on Fig.~\ref{fig:Fig4_supporting_2}e with the blue curve. We then compare this with the typical trajectory length an electron undertakes traveling from the S electrode to the barrier. To calculate this, we have to determine the diffusion coefficient in the wire, again using the reference segment. We have used the Einstein relation: 

\begin{equation}\label{eq:einstein}
D=\frac{\sigma}{e^2\rho(E_\mathrm{F})},
\end{equation}
where $\rho(E_{\mathrm{F}})$ is the density of states (3D, parabolic dispersion):
\begin{equation}\label{eq:dos}
\rho(E_\mathrm{F})=\frac{m^{*2}}{\pi^2\hbar^3}v_\mathrm{F},
\end{equation}
where $\hbar$ is the reduced Plank's constant. The diffusion coefficient is plotted in Fig.~\ref{fig:Fig4_supporting_2}d. Finally using the equation of 1D diffusion:
\begin{equation}\label{eq:dwell}
L=\sqrt{D\tau_\mathrm{{dw}}}.
\end{equation}
The estimated effective trajectory length is shown in Fig.~\ref{fig:Fig4_supporting_2}e with red:

\begin{equation}\label{eq:path}
L_{\mathrm{tr}}=\frac{L^2}{D}v_\mathrm{F}.
\end{equation}
Here we have introduced $\tau_{\mathrm{dw}}=L_{\mathrm{tr}}/v_{\mathrm{F}}$, the dwell time. The result is in qualitative agreement with the length obtained from the ABS data, which taken into all the unknown parameters is quite reasonable. 

Finally we have also extracted the coherence length inside the wire using: 

\begin{equation}\label{eq:xi_diffusive}
\xi=\sqrt{\frac{\hbar D}{\vert\Delta\vert }}
\end{equation}

This is shown in Fig.~\ref{fig:Fig4_supporting_2}f. which shows an increase towards higher gate voltages and is comparable to the junction length. This can lead to an increase of the bound state energies (induced gap size) and can explain the transition towards the short junction limit at high gate voltages.

\bibliography{Suppinfo}